\documentclass[11pt]{article}

\usepackage{graphicx}
\usepackage{hyperref}
\usepackage{amsmath}
\usepackage[dvipsnames]{xcolor}

\usepackage[numbers]{natbib}

\newcommand{\Mstrict}{{Mk-1}}
\newcommand{\Mrelaxed}{{Mk-8}}
\newcommand{\ncanidfossils}{{116}}

\begin{document}

\bigskip
\medskip
\begin{center}

\noindent{\Large \bf Bayesian phylogenetic estimation of fossil ages} 

\bigskip

\noindent {\normalsize \sc  Alexei J.\ Drummond$^{1,2,3}$ and Tanja Stadler$^{3,4}$}\\
\noindent {\small \it 
$^1$Centre for Computational Evolution, University of Auckland, Auckland, New Zealand;\\
$^2$Department of Computer Science, University of Auckland, Auckland, 1010, New Zealand;\\
$^3$Department of Biosystems Science \& Engineering, Eidgen\"{o}ssische Technische Hochschule Z\"{u}rich, 4058 Basel, Switzerland;\\
$^4$Swiss Institute of Bioinformatics (SIB), Switzerland.}
\end{center}
\medskip
\noindent{\bf Corresponding author:} Alexei J. Drummond, Centre for Computational Evolution, University of 
Auckland, Auckland, New Zealand; E-mail: alexei@cs.auckland.ac.nz\\

% 250 words
\abstract{
Recent advances have allowed for both morphological fossil evidence and molecular sequences to be integrated into a single combined inference of divergence dates under the rule of Bayesian probability. 
In particular the fossilized birth-death tree prior and the Lewis-Mk model of discrete morphological evolution allow for the estimation of both divergence times and phylogenetic relationships between fossil and extant taxa. 
We exploit this statistical framework to investigate the internal consistency of these models by producing phylogenetic estimates of the age of each fossil in turn, within two rich and well-characterized data sets of fossil and extant species (penguins and canids).
We find that the estimation accuracy of fossil ages is generally high with credible intervals seldom excluding the true age and median relative error in the two data sets of 5.7\% and 13.2\% respectively. The median relative standard error (RSD) was 9.2\% and 7.2\% respectively, suggesting good precision, although with some outliers. 
In fact in the two data sets we analyze the phylogenetic estimates of fossil age is on average  $< 2$ My from the midpoint age of the geological strata from which it was excavated. 
The high level of internal consistency found in our analyses suggests that the Bayesian statistical model employed is an adequate fit for both the geological and morphological data, and provides evidence from real data that the framework used can accurately model the evolution of discrete morphological traits coded from fossil and extant taxa. 
We anticipate that this approach will have diverse applications beyond divergence time dating, including dating fossils that are temporally unconstrained, testing of the ``morphological clock'', 
and for uncovering potential model misspecification and/or data errors when controversial phylogenetic hypotheses are obtained based on combined divergence dating analyses.
}

\section*{Introduction}

% Papers
% http://www.ncbi.nlm.nih.gov/pmc/articles/PMC4033271/ Beyond fossil calibrations: realities of molecular clock practices in evolutionary biology
% http://journals.plos.org/plosone/article?id=10.1371/journal.pone.0066245 A Simple Method for Estimating Informative Node Age Priors for the Fossil Calibration of Molecular Divergence Time Analyses
% http://www.ncbi.nlm.nih.gov/pmc/articles/PMC3108556/  Bayesian Phylogenetic Method to Estimate Unknown Sequence Ages
% http://www.pnas.org/content/112/16/4897 Canids

Contention between palaeontologists and molecular biologists over which data provides the most accurate inferences about evolutionary history has previously fostered an adversarial relationship between the two fields \cite{Donoghue2007}. 
Although there has indeed been much controversy surrounding apparent discrepancies between palaeontological and molecular phylogenetic inferences \cite{BromhamPenny2003} it is also clear that fossil and molecular data both produce broadly concordant views of evolutionary history \cite{BentonAyala2003}. 
%Indeed one of the first examples of applying a macroevolutionary phylogenetic model to perform inference directly on fossil occurrence data in fact demonstrated concordance between inferences from  molecular data and palaeontological data in primates \cite{tavare2002using}. 
% AJD: The following two sentences are inflammatory to palaeontologists
%Nevertheless there have been only few attempts to apply phylogenetic reasoning to palaeontological questions. 
%Models of molecular evolution have been refined over many decades \cite{Felsenstein2004,Yang:2006yu}. 
The continual improvement of models and methods for statistical phylogenetic inference from molecular sequence data is well documented \cite{Felsenstein2004,Yang2014}, and in recent years it is arguably the case that molecular phylogenetics has taken primacy over the fossil record in providing a timescale for evolutionary history \cite{Donoghue2007}. 
Nevertheless molecular phylogenetic inference of evolutionary timescales relies critically on calibration by the fossil record \cite{Donoghue2007}.

%In particular probabilistic model based approaches to biogeography, ancestral state reconstruction, and rates of morphological character change have been common in recent years. Nevertheless there has been a call by some palaeontologists for 
%The earlier literature is awash with research papers that include fossils in phylogenies, utilize a phylogenetic bracket to constrain inferences of soft tissue or physiology, or use trees to reconstruct biogeography.

Traditionally the practice has been to use one or more fossils as ``node calibrations'' by associating their geologically-derived age to a particular divergence in a molecular phylogeny. The age of the fossil is determined either by radiometric aging of strata above and/or below the fossil, or more commonly by biostratigraphy. 
The difficulty lies in determining the appropriate ancestral divergence in the molecular phylogeny to associate the fossil with and the details of how this should be achieved within a full statistical inference framework \cite{Ho2009, Heled2012, HeledDrummond2015}.
Once achieved, node calibration confers age estimates to the remaining ancestral divergences in the phylogenetic tree by the assumption of a strict or relaxed molecular clock \cite{Thorne1998,thorne2005,yang2006,Drummond2006,drummond2010}. 

It may be less widely appreciated by molecular evolutionary biologists that the statistical phylogenetic revolution in molecular evolution has also been mirrored in the increasing application of statistical phylogenetic reasoning in macroevolutionary and systematic studies of the fossil record \cite{Foote1996,huelsenbeck1997maximum, tavare2002using, WagnerMarcot2013}. 
%This includes applying phylogenetic reasoning to questions of biogeography, character evolution, and the completeness of the fossil record.
Here we extend this tradition of applying phylogenetic reasoning to the fossil record by focusing on the question of what phylogenetic inference techniques can tell us about the age of a fossil, based solely on its morphological characteristics and through them, its phylogenetic and temporal relationships with a set of reference fossils.

The phylogenetic estimation of the age of a taxon based on its molecular sequence has been previously described \cite{drummond2002computational,shapiro2011bayesian} and applied to both ancient subfossil remains and rapidly evolving viral taxa. 
For example, this technique has been successfully employed to estimate the age of human subfossil remains based on an ancient mitochondrial genome sequence \cite{meyer2014mitochondrial}. 
The same technique has also been used to estimate the age of viral samples based on molecular sequence data (e.g. \cite{gray2013evolutionary}).

We extend this approach into the realm of discrete morphological evolution by presenting a statistical model of evolution that generates an expectation on the distribution of fossils, their morphological characters. 
This model has been previously presented in the context of divergence time dating \cite{gavr2014,gavryushkina2015bayesian} and \cite{zhang2016}. 

%TSrev I moved this section up
%AJD revision
%\subsection*{Rates of evolution and relaxed morphological clocks}
In order to use discrete morphological comparative data to estimate fossil ages, it is necessary to assume a (relaxed) morphological clock. There is a long history of the study of the evolutionary rates of phenotypic characters \cite{Simpson1944,Haldane1949,Simpson1953,Gingerich1983,Gingerich1993}, going at least back to Darwin's Origin of Species \cite{Darwin1859}. Darwin noted that {\it ``Species of different genera and classes have not changed at the same rate''} and illustrated this point with examples of ``living fossils'' such as the Silurian mollusc {\it Lingula} \cite{Darwin1859}. However in the same chapter Darwin goes on to say {\it ``In members of the same class the average amount of change, during long and equal periods of time, may, perhaps, be nearly the same''}. Nevertheless, phenotypic evolution has more typically been characterized as not evolving in a clock-like manner, especially when compared to molecular evolution \cite{Kimura1983}. While there are many examples of extremely slow and fast rates of phenotypic evolution in the literature, we would argue that this is also true for molecular rates. We are not aware of a comprehensive and systematic comparison of variation in evolutionary rates at the phenotypic and molecular levels. Regardless, for the data sets that we analyze, we adopt the point of view that variation in the rate of phenotypic evolution across the phylogeny can be accommodated with a relaxed morphological clock.

Our approach is distinct from alternative divergence time dating approaches in that it provides an explicit treatment of the temporal information contained in fossil remains, whether or not related molecular sequence data is available. 
This leads to an estimate of the age of the most recent common ancestor of a group of fossil and extant taxa. A key difference between this approach and earlier approaches to tip-calibrated ``total-evidence'' dating \cite{Ronq2012} is the admission of a probability that each fossil taxon may represent a sampled ancestor of one or more taxa  in the tree \cite{gavr2014}.
We exploit this framework to attempt the estimation of the age of individual fossils based solely on morphological data and their phylogenetic affinities to related taxa of known age. 
The method is applied to two rich and well-characterized morphological data sets: (i) 
%TS how many extant penguins?
19 extant penguins and 36 fossil relatives \cite{ksepka2010,ksepka2012}, (ii) a sample of nine extant canids and \ncanidfossils{} fossil relatives \cite{Slater2015}.

%http://earth.geology.yale.edu/~ajs/1993/11.1993.17Gingerich.pdf
%https://books.google.co.nz/books?id=olIoSumPevYC&printsec=frontcover#v=onepage&q&f=false

%%%

\section*{Methods}
%TSrev
Gavryushkina {\it et al} \cite{gavryushkina2015bayesian} described a ``total-evidence'' approach implemented in BEAST2 \citep{Beast2} for phylogenetic estimation of time-trees that employs both morphological data from fossils and extant data and molecular sequence data as equal partners under the rule of probability for estimating a time-tree.  An equivalent method \citep{zhang2016} is introduced within MrBayes \cite{ronquist2012mrbayes}. 
The model of time-tree phylogeny employed is the so-called fossilized birth-death process \cite{Heath2014}, which forms a prior probability distribution on the space of sampled-ancestor trees \cite{gavr2014,Gavr2013}.

We extend the approach in \cite{gavryushkina2015bayesian}  further by investigating the consistency between the phylogenetic estimate of the age of a fossil and the corresponding fossil age range determined by geological and biostratigraphic evidence. 
%Previous study described how a set of fossils with discrete morphological characters could be used to estimate a time-tree \cite{gavryushkina2015bayesian,zhang2016}. 
%Here we additionally allow for one or more of the fossils to have broad uninformative priors on their age. 
This allows for the age of some of the fossils to be estimated solely based on their morphological characters and the phylogenetic affinities of their morphology to other fossils with known ages in the time-tree. 
We refer to this as the phylogenetic estimate of the fossil's age. 
In phylogenetically estimating the age of each of the fossils in turn, two questions can be answered: (i) How much information about an individual fossil's age is available from phylogenetic analysis of morphological data alone, and (ii) What is the level of phylogenetic evidence in support of the palaeontological age range for a fossil?
These two questions are investigated using two morphological data sets, one of 36 fossil penguins and their extant relatives \cite{ksepka2010,ksepka2012,gavryushkina2015bayesian}, and one of \ncanidfossils{} canid fossils and their extant relatives \cite{Slater2015}.

\subsection*{Phylogenetic estimates of the ages of penguin fossils}

%\alexei{%
We used a data set originally published by \cite{ksepka2012} consisting of morphological data from fossil and living penguin 
species. 
We used the same subset of the morphological data as in \cite{gavryushkina2015bayesian} but we did not use the molecular sequence data from the living species. 
The morphological data matrix we used contains 36 fossil species, 19 extant species and 202 characters (ranging from binary to k=7). 
The majority of these characters (\textgreater 95\%) have fewer than four states and 48 of the binary characters were encoded as presence/absence. 
The fossil age intervals had median values ranging from 5.55 to 61.05 Myr.
As did \cite{gavryushkina2015bayesian}, we treat 34 characters that were ordered in \cite{ksepka2012} as unordered. 
See \cite{gavryushkina2015bayesian} for further details of data selection.
%}%

For each of the 36 penguin fossils in turn we performed a separate Bayesian phylogenetic analysis in which the focal fossil's palaeontological age constraints were replaced by 
the fossilized birth-death process prior, and thus we obtained a phylogenetic estimate of the fossil's age.  

\subsection*{Phylogenetic estimates of the ages of Canids fossils}

%\alexei{%
The second data set that we investigated was a morphological data matrix of 125 canid species \cite[9 extant and \ncanidfossils{} fossil;][]{Slater2015} with 122 characters (ranging from binary to k=5).
%}%
The 9 extant species represent about 25\% of the extant canid species and include representatives of four genera (6 {\em Canis}, 1 {\em Cuon}, 1 {\em Lycaon}, 1 {\em Urocyon}) and both tribes (8 Canini, 1 Vulpini). 
%\alexei{%
We had stratigraphic ranges based on palaeontological data for all \ncanidfossils{} fossils (Graham Slater, pers. comms). 
%}%

As with the penguin data set we  performed an analysis for each of the \ncanidfossils{} canid fossils in turn. 
 Unlike the original study \cite{Slater2015} we did not apply any other constraints or priors on ancestral divergence times beyond the ages of the fossils. 

\subsection*{Phylogenetic analyses}

%TSrev
Given the morphological data $D$ and a stratigraphic age range for each fossil $a = ((l_1,u_1),(l_2, u_2),\ldots,(l_n,u_n))$ (with $l_i$ being the lower age bound for fossil $i$, and $u_i$ being the upper age bound for fossil $i$), we sample phylogenetic trees with the fossils being tips or sampled ancestors, and each fossil having a specified age  in the phylogenetic tree within its stratigraphic age range. The parameters of the fossilized birth-death model are summarized in $\eta$, and the parameters  of the model for morphological character evolution are summarized in $\theta$. More formally, we sample from,
$$P[\mathcal{T},\eta, \theta | D, a] = P[D|\mathcal{T},\theta] P[\mathcal{T},a|\eta] P[\eta] P[\theta]/P[D,a],$$
where $ P[\mathcal{T},a|\eta] =  P[\mathcal{T}|\eta] $ if each fossil age is within its stratigraphic age range specified in $a$, and $ P[\mathcal{T},a|\eta] =0$ else.
When we replaced the focal  palaeontological age constraints of fossil $i$  by 
the fossilized birth-death process prior, we simply set $l_i=0$ and $u_i=T$ (where 0 is present time and $T$ is the total height of the tree) to estimate the phylogenetic age of the focal fossil.

The fossilized birth-death model is defined by the following parameters $\eta=(T,d,r,s)$:  the time of the start of the process $T$ prior to present time 0, the net diversification rate ($d$ = speciation rate - extinction rate), the turnover $r$ (= extinction rate / speciation rate) and the sampling probability with which a fossil is observed $s$ (=sampling rate / (extinction rate + sampling rate)).
  %%%
  
Following \cite{gavryushkina2015bayesian}, we apply the Lewis Mk model 
\cite{Lewis2001} for discrete morphological character evolution, which assumes a character can take $k$ states and the transition rates from one state to another are equal for all states. 

%TSrev
We applied two phylogenetic models to the penguin data set, \Mstrict{} and \Mrelaxed{}, and we applied \Mstrict{} to the canid dataset. \Mstrict{} assumed a strict morphological clock ($\mu$) and no gamma-distributed rate heterogeneity among sites  \cite{Lewis2001}. 
Model \Mrelaxed{} \citep{gavryushkina2015bayesian} partitioned the alignment into  partitions (six for the penguins), with the $i$'th partition containing all characters that had $k=i+1$ character states across the sampled taxa. 
Model \Mrelaxed{} also uses a uncorrelated lognormally-distributed relaxed molecular clock \cite{Drummond2006} with parameters $\mu$ and $S$ for the mean rate and log standard deviation of the rates, and an additional parameter $\alpha$ governing the shape for gamma-distributed rate variation across sites \cite{yang:1994ma}. 
The prior distribution for $\alpha$ was uniform in the interval $(0,10)$. 

We followed \cite{gavryushkina2015bayesian} in having a broad LogNormal($M=-5.5, S=2$) prior on $\mu$ for all analyses and a Gamma($\alpha=0.5396, \beta=0.3819$) prior on $S$ for the relaxed clock analyses.
For the penguin analyses, the parameters of the fossilized birth-death model tree prior  were specified as described in the section ``Computing the phylogenetic evidence for an age range''. 
Since we did not perform Bayes factor (BF) analyses for the canid dataset, we used the standard parametrisation $\eta=(T,d,r,s)$, with the following priors: uniform prior from in the interval $(0, 120)$ million years for origin $T$, LogNormal($M=-3.5, S=1.5$) prior for diversification rate $d$, unit uniform prior $(0,1)$ for turnover $r$ and sampling proportion $s$.

\subsection*{Computing the phylogenetic evidence for an age range}

The Bayes factor ($BF$) computes the evidence for one hypothesis ($H_1$) over another ($H_2$) as the ratio of the marginal probability of the data under each of the two hypotheses and a model $M$, 

\begin{equation}
BF = \frac{p(D|H_1,M)}{p(D|H_2,M)} = \frac{p(H_1|D,M)}{p(H_2|D,M)}\frac{p(H_2|M)}{p(H_1|M)}.
\end{equation}

We are interested in computing the Bayes factor that quantifies the amount of phylogenetic evidence in support of the palaeontological age range for each fossil. In this case $H_1$ is the hypothesis that the true fossil age is within the given palaeontological age range, and $H_2$ is the alternative hypothesis that the true fossil age is outside the palaeontological range. A BF $>>$ 1 indicates strong support for $H_1$, given the model $M$ is appropriate for the considered data.

%, but within the broader interval of $(0,T_\text{max})$ Mya. 
The model $M$ consists of two parts, $M=(M_\mathcal{T},M_m)$. The model $M_\mathcal{T}$ specifies the tree generation process  giving rise to the number of observed samples and sampling times. The model $M_m$ specifies the morphological evolution along the tree giving rise to the morphological characters for the samples. The data $D=(D_\mathcal{T},D_m)$ is the number of samples together with the sampling times ($D_\mathcal{T}$) and  the morphological characters for each sample ($D_m$). 

For caclulating the Bayes factor, the probabilities $p(H_1|D,M)$ and $p(H_2|D,M)$ are obtained directly from the MCMC output.
It remains to calculate the probabilities $p(H_1|M)$ and $p(H_2|M)$.
Since $H_1$ and $H_2$ are independent of $M_m$, we have $$\frac{p(H_2|M)}{p(H_1|M)}=\frac{p(H_2|M_\mathcal{T})}{p(H_1|M_\mathcal{T})}=\frac{1-p(H_1|M_\mathcal{T})}{p(H_1|M_\mathcal{T})}.$$

One way to determine $p(H_1|M_\mathcal{T})$ would be to simulate trees under the model $M_\mathcal{T}$ and record the fraction of sampling times  within a given palaeontological age range. However, such a simulation approach turns out to be very time-consuming, and the procedure below provides a much faster evaluation of $p(H_1|M_\mathcal{T})$.

%TSrev
We   derive some analytic results for evaluating $p(H_1|M_\mathcal{T})$. 
The model $M_\mathcal{T}$ is the fossilized birth-death process with priors on its parameters $\eta=(T,d,r,s)$. 
%Using the parameterization in \cite{gavryushkina2015bayesian}, the parameters are the time of the start of the process $T$, the net diversification rate $d$ (= speciation rate - extinction rate), the turnover $r$ (= extinction rate / speciation rate) and the sampling probability $s$ (=sampling rate / (extinction rate + sampling rate)).
We derive the probability density of sampling a fossil at time $t$ in the past, given the model $M_\mathcal{T}$. This probability density will allow us to directly determine $p(H_1|M_\mathcal{T})$.

For a given $T$, $d$, $r$ and $s$, the probability density of sampling a fossil at time $t$, given the process does not go extinct for time $T$, is,
$$p(t|T,d,r,s) = \frac{1}{1-p_0(T;d,r)} \sum_{k=1}^\infty k \psi p_k(T-t;d,r) (1-p_0(t;d,r)^k)$$
with $\psi=\frac{s}{1-s} \frac{rd}{1-r}$ being the sampling rate, and $p_i(t;d,r)$ being the probability of a single lineage producing $i$ surviving lineages at time $t$.
The equation above calculates the required probability and the left term in that probability conditions on survival of the process ($1-p_0(T;d,r)$). Then we calculate the probability to have $k$ lineages at time $t$ before the present $( p_k(T-t;d,r))$, multiply by the sampling rate $k \psi$, and weight by the probability that at least one lineage of the $k$ lineages survives to the present $ (1-p_0(t;d,r)^k)$. This expression is then summed over $k=1,\ldots, \infty$.

We simplify, using the equations for $p_i(t;d,r)$ given in \cite{kendall1948}, to obtain,
\begin{eqnarray*}
p(t|T,d,r,s) &=& \frac{1}{1-p_0(T;d,r)}  \psi p_1(T-t|d,r) \\ & & \times \left[  \frac{1}{(1-p_0(T-t|d,r)/r)^2} - \frac{p_0(t|d,r)}{(1-p_0(t|d,r) p_0(T-t|d,r)/r)^2}  \right]\\
&=&  \psi  \left[ \frac{e^{d(T-t)}}{1-p_0(T;d,r)} - \frac{p_0(T|d,r)-p_0(T-t|d,r)}{(1-p_0(T-t|d,r))(1-p_0(t|d,r))} \right],
\end{eqnarray*}
with $p_0(t|d,r) = \frac{1-e^{-d*t}}{1/r - e^{-d*t}}$.

Next, we need to evaluate $p(t|M_\mathcal{T}) = \int_{T,d,r,s} p(t|T,d,r,s) p(T) p(d) p(r) p(s)$ with $p(T), p(d), p(r)$ and $p(s)$ being the prior distributions for the parameters. This is done by sampling parameters from the prior distributions, and then evaluating $p(t|T,d,r,s)$. 

\begin{figure}
\includegraphics[width=12cm]{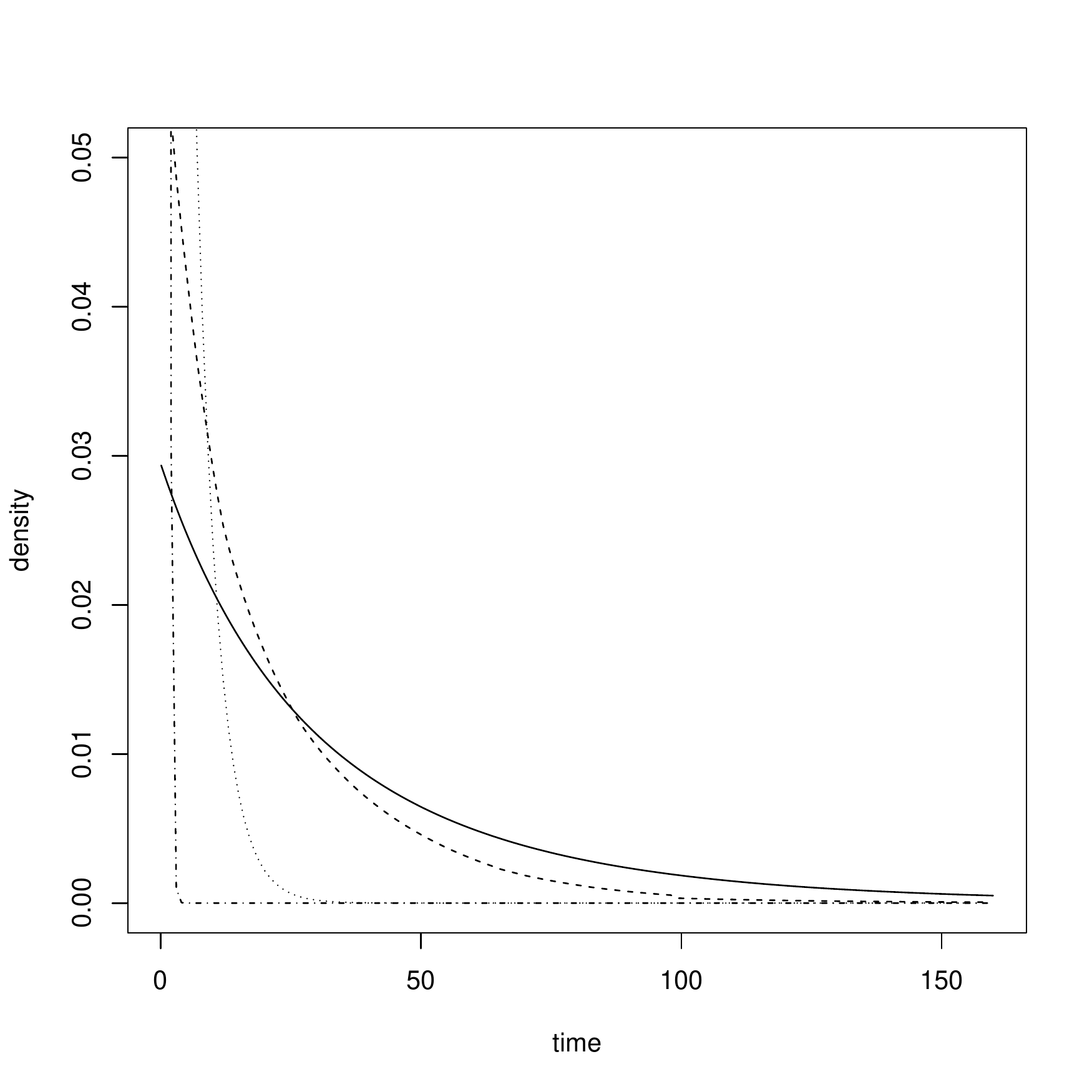}
\caption{\label{hist8_older}Probability density for the sampling times under the fossilized birth-death process. The dot-dashed line uses priors on the parameters as in \cite{gavryushkina2015bayesian}. 
The solid line uses the new prior with implicit assumptions on $T$ and $s$, the dashed line results from only assuming the implicit prior on $T$, the dotted line results from only assuming the  implicit prior on $s$. 
(Since the dashed, dotted and dot-dashed lines are governed by rare parameter combinations leading to huge trees and huge sample sizes, these lines are very sensitive to drawing another rare event, and thus need to be taken with some caution).}
\label{Fig:Prior}
\end{figure}

We determined $p(t|M_\mathcal{T})$ for the prior distributions as in \cite{gavryushkina2015bayesian}, 
$$T: Unif(0,160), d: lognorm(-3.5,1.5), r: Unif(0,1), s: Unif(0,1).$$ This prior specification leads to a distribution of sampling time with almost all probability mass close to the present (Figure \ref{Fig:Prior}, dot-dashed line). Thus, essentially $p(H_1|M_\mathcal{T})=0$  which leads to a huge Bayes Factor. This means we always reject $H_2$, but not because we necessarily agree with the palaeontological age range, but because our model has no prior weight for the palaeontological age range.

Inspection of our prior identifies two problems: (i) If we draw a large $T$ and large $d$, we obtain very large trees with arbitrarily many species close to the present, thus we have most of the sampling times close to the present, (ii) If we draw $r$ and $s$ close to 1, then we obtain a very large per-lineage  sampling rate $\psi=\frac{s}{1-s} \frac{rd}{1-r}$. 
Thus these parameter combinations govern the probability density curve and cause again most prior weight to be close to the present.

We therefore assumed new prior distributions. The net diversification rate $d: LogNormal($M=-3.5,S=0.5$)$ was chosen with a smaller standard deviation which avoids too much weight on very fast growing trees. The turnover $r: Uniform(0,1)$ was set as before. 

For $s$ we assume an implicit prior: we assume $LogNormal(-2,1)$ for $\psi$, and $$s=\psi/(\mu+\psi)$$ (with extinction rate $\mu = r d /(1-r)$). This avoids very high sampling rates.

For $T$, we also assume an implicit prior. We assume a uniform distribution on $[1,100]$ for the number of present day species, $N$. In expectation, we have $N=  \frac{e^{d T}}{1-p_0(T)}$ species after time $T$. This leads to $$T=\log((1-r)N +r)/d.$$
Overall, this prior produces a sampling time distribution where old sampling times have a non-negligable weight (Figure \ref{Fig:Prior}, solid line). 
The choice of an implicit prior for both $T$ and $s$ was important: only specifying the implicit prior on $T$ yields the dashed line in Figure \ref{Fig:Prior}, while only specifying the implicit prior on $s$ yields the dotted line in Figure \ref{Fig:Prior}.
We used this new prior for our  analyses and the Bayes Factor calculation.

Changing to our new prior has immense impact on the Bayes factor analysis, but in our case has a minor effect on the posterior distribution of trees / parameters compared to using the prior in  \cite{gavryushkina2015bayesian}.
This investigation of the prior distribution on trees and sampling times highlights that whenever using Bayes factors to test a hypothesis, we have to first investigate what our prior on the hypothesis is. In our example, the  prior from \cite{gavryushkina2015bayesian} seemed reasonable for the parameters specified, however this prior puts a negligible weight on hypothesis $H_1$ for older fossils.

We want to note that the stepping stone sampling approach \cite{Xie2011}
 to calculate Bayes factors would not have been directly applicable in our case:  In stepping stone, sampling the $D_\mathcal{T}$ is treated as part of the model, not part of the data. However, using a birth-death model, the sampling times are part of the data. The approach is valid when choosing a  coalescent tree prior, as in that case sampling times are conditioned upon (and thus can be seen as part of the model assumptions) rather than being modelled (and thus are a realisation of the model which means they are data). It is not clear if the stepping stone approach can be directly applied for models with number of tips being part of the data.
In general, even if stepping stone approaches are appropriate, we recommend inspection of $P(H_1|M)$ to ensure that the prior on the hypothesis to be tested is sensible. Such an investigation reveals if the cause of a high (or low) Bayes factor is due to the prior or due to signal in the data.

\section*{Results}

\subsection*{Penguins conform well to a morphological clock}
Although \Mstrict{} is a very simple model, the phylogenetic estimates of the ages of the penguin fossils were remarkably consistent with their palaeontological age ranges. 
Figure \ref{fig:phyloAgeVsGeoAge}a plots the geological age and range against the phylogenetic estimates of fossil age. The points in this plot have $R^2 = 0.903$. 
The median error (where the {\it error} is the difference between the phylogenetic median and the geological median) is 1.96 Myr. The median relative error (where the {\it relative error} is the {\it error} divided by the geological median) was 5.7\% and the median relative standard deviation (RSD; defined as the standard deviation of the marginal posterior divided by the posterior median estimate) was 9.2\%.
A summary of the individual estimates are tabulated in Table \ref{fossilTable1}.

As judged by Bayes factors, only one fossil exhibited strong evidence (i.e. $\text{log BF} < -3.0$) that the phylogenetic estimate of fossil age was inconsistent with the geological age range. The log BF for {\em Paraptenodytes antarcticus} was -3.4. In fact the majority of the fossils (23/36 = 64\%) had strong positive evidence for the geological age range (i.e. $\text{log BF} > 3.0$).
Likewise, if we consider only the posterior probability that the fossil is in the geological age range, then three of the 36 fossils has a posterior probability $< 0.05$, suggesting low posterior support for the phylogenetic estimate of fossil age being within the palaeontological age range.
These three fossils were {\em Madrynornis mirandus}, {\em Paraptenodytes antarcticus} and {\em Sphenicus muizoni} with posterior probabilities that the phylogenetic estimate of fossil age is in the palaeontological range of 0.007, 0.001 and 0.001, respectively. 
All other fossils have posterior probabilities of $> 0.05$ of their age being in the palaeontological range. 
It is worth noting that the absolute discrepancy in the ages are still quite moderate for the three fossils with low posterior probabilities, with {\em M. mirandus}: 6.3 Myr vs 10 Myr (phylogenetic estimate of fossil age versus palaeontological age), {\em P. antarcticus}: 29.9 vs 22, {\em S. muizoni}: 5.2 vs 9.1. 
The small posterior probabilities are partially caused in these cases because the corresponding palaeontological age range is narrow, apparently suggesting very precise geological knowledge of the ages of these three fossils.

\subsection*{Relaxing the clock, site partitions, rate variation among sites}

\begin{figure}
\includegraphics{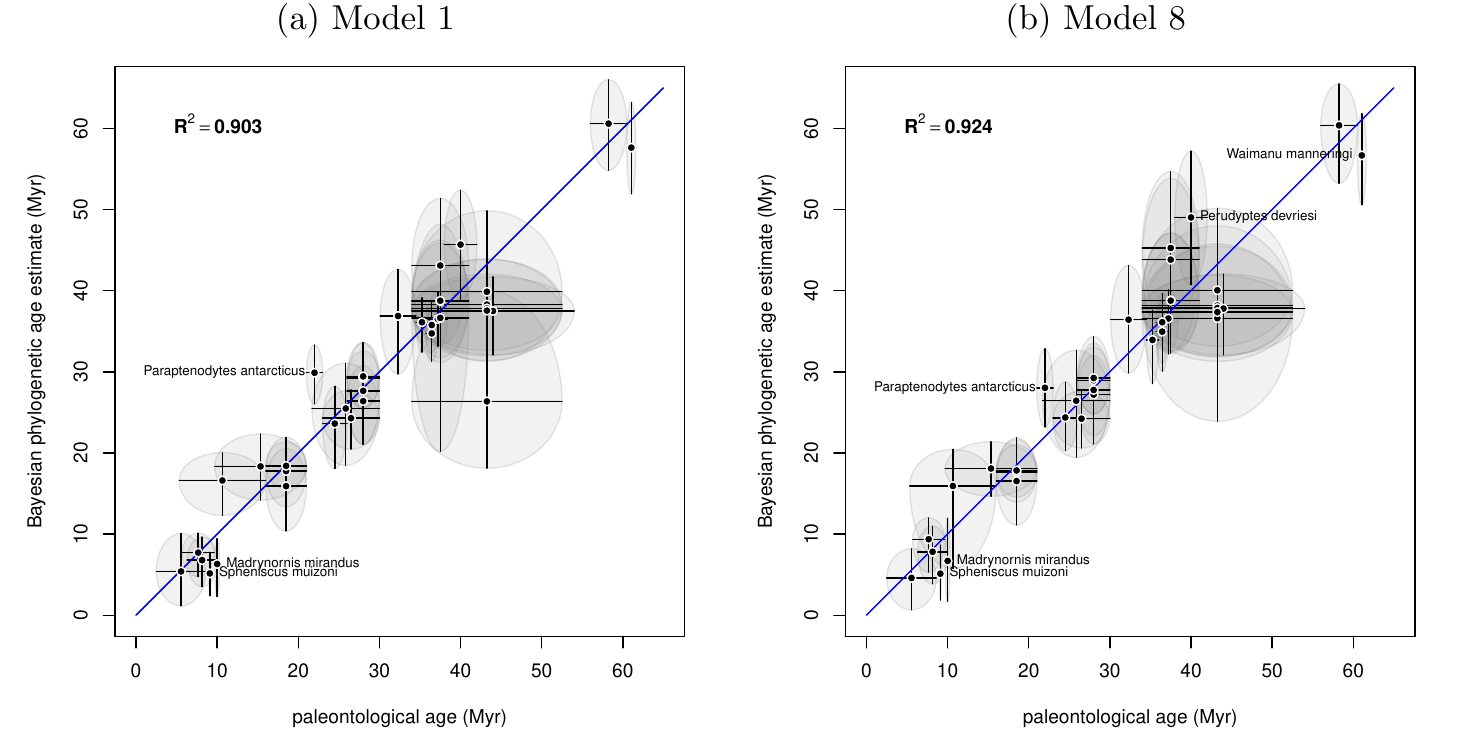}
\caption{\label{fig:phyloAgeVsGeoAge}
The Bayesian phylogenetic estimate of fossil age (median of marginal posterior) for each of the 36 penguin fossils plotted against their palaeontological age estimates, under two alternative site and molecular clock models. 
The palaeontological age estimates are represented by the mid-point of the range and the upper and lower limits. 
The Bayesian estimates are represented by the median of the marginal posterior distribution and the upper and lower limits of the 95\% HPD interval. 
The blue line shows the $x=y$. If the vertical line doesn't cross $x=y$, then the midpoint of the geological range is not in the phylogenetic 95\% HPD. 
If the horizontal line doesn't cross $x=y$, then the median phylogenetic estimate is not contained in the palaeontological age range. 
The three labelled fossils have posterior probability of less than 0.05 for their age being within the palaeontological age interval.
%TS do these three fossils not intersect in horizontal AND vertical?
}
\end{figure}

\Mrelaxed{} was the best-fitting model for the penguin data set according to the analysis of \cite{gavryushkina2015bayesian}. 
As with \Mstrict{} this model produced phylogenetic estimates of fossil age that were very concordant with the geological age ranges of the fossils (Figure \ref{fig:phyloAgeVsGeoAge}b), with an overall $R^2 = 0.924$. The median error was 2.05 Myr across all 36 fossils. 
In this analysis none of the fossils exhibited any evidence (i.e. $\text{log BF} < 0.0$) that the phylogenetic estimate of fossil age was inconsistent with the geological age range. 
However if we consider the posterior probability that the fossil is in the geological age range then five of the 36 fossils had a posterior probability $< 0.05$ for \Mrelaxed{}, suggesting low posterior support for the phylogenetic estimate of fossil age being within the palaeontological age range. 
These five fossils were {\em Madrynornis mirandus}, {\em Paraptenodytes antarcticus}, {\em Perudyptes devriesi}, {\em Sphenicus muizoni} and {\em Waimanu manneringi} with posterior probabilities that the phylogenetic estimate of fossil age is in the palaeontological range of 0.035, 0.018, 0.046, 0.004, 0.037 respectively. 
All other fossils have posterior probabilities of $> 0.05$ of their age being in the palaeontological range. 
Again the absolute discrepancy in the ages are quite moderate for the five fossils with low posterior probabilities, with {\em M. mirandus}: 6.7 Myr vs 10 Myr (phylogenetic estimate of fossil age versus palaeontological age), {\em P. antarcticus}: 28.0 vs 22, {\em P. devriesi}: 49.0 vs 40, {\em S. muizoni}: 5.1 vs 9.1 and {\em W. manneringi}: 56.7 vs 61.05. 
A summary of all the individual estimates are tabulated in Table \ref{fossilTable8}. 
The individual marginal posterior distributions of phylogenetic estimates of fossil age under \Mrelaxed{}, and the corresponding geological range are shown in Figures \ref{hist8_younger} and \ref{hist8_older}.

\begin{figure}
\includegraphics[width=5in]{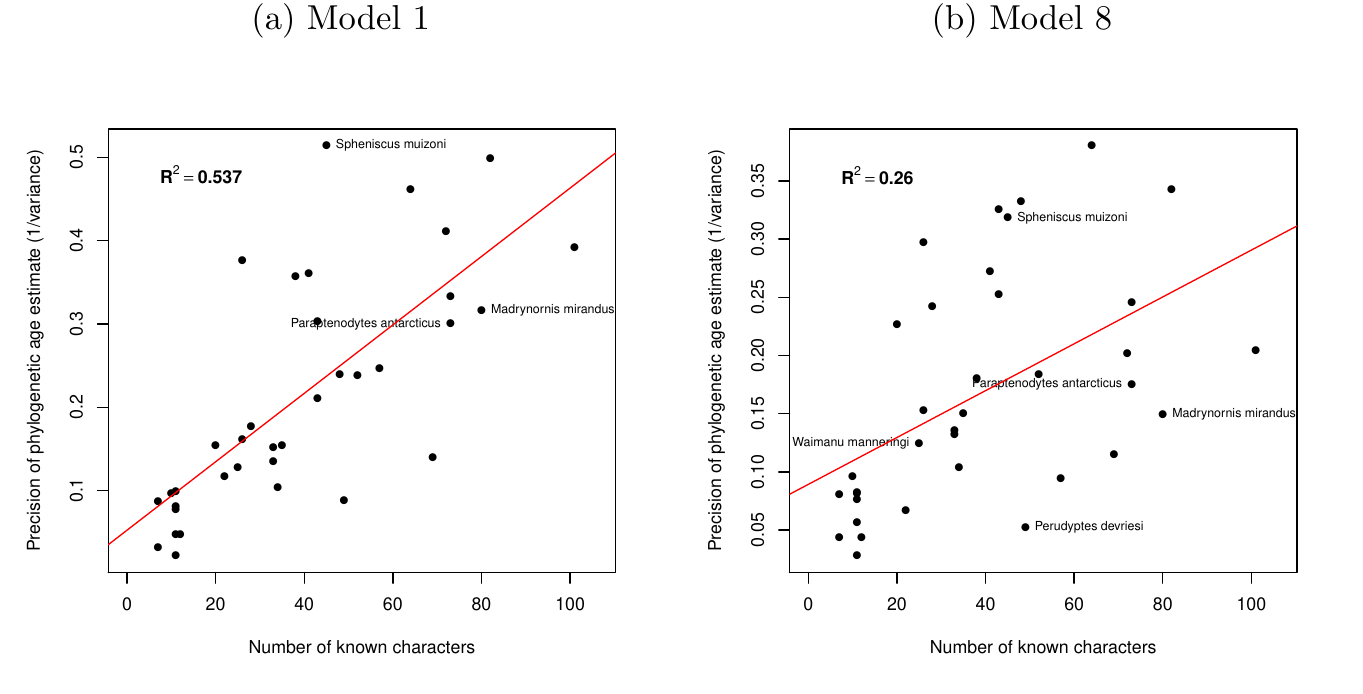}
\caption{\label{fig:precisionVsnumberCharacters} A plot of the number of non-ambiguous morphological characters for the penguin taxon against the precision of the phylogenetic estimate of corresponding fossil age for (a) \Mstrict{} and (b) \Mrelaxed{} (i.e. the precision is 1/variance in the marginal posterior distribution of the age).}
\end{figure}

\begin{figure}
\includegraphics[width=12cm]{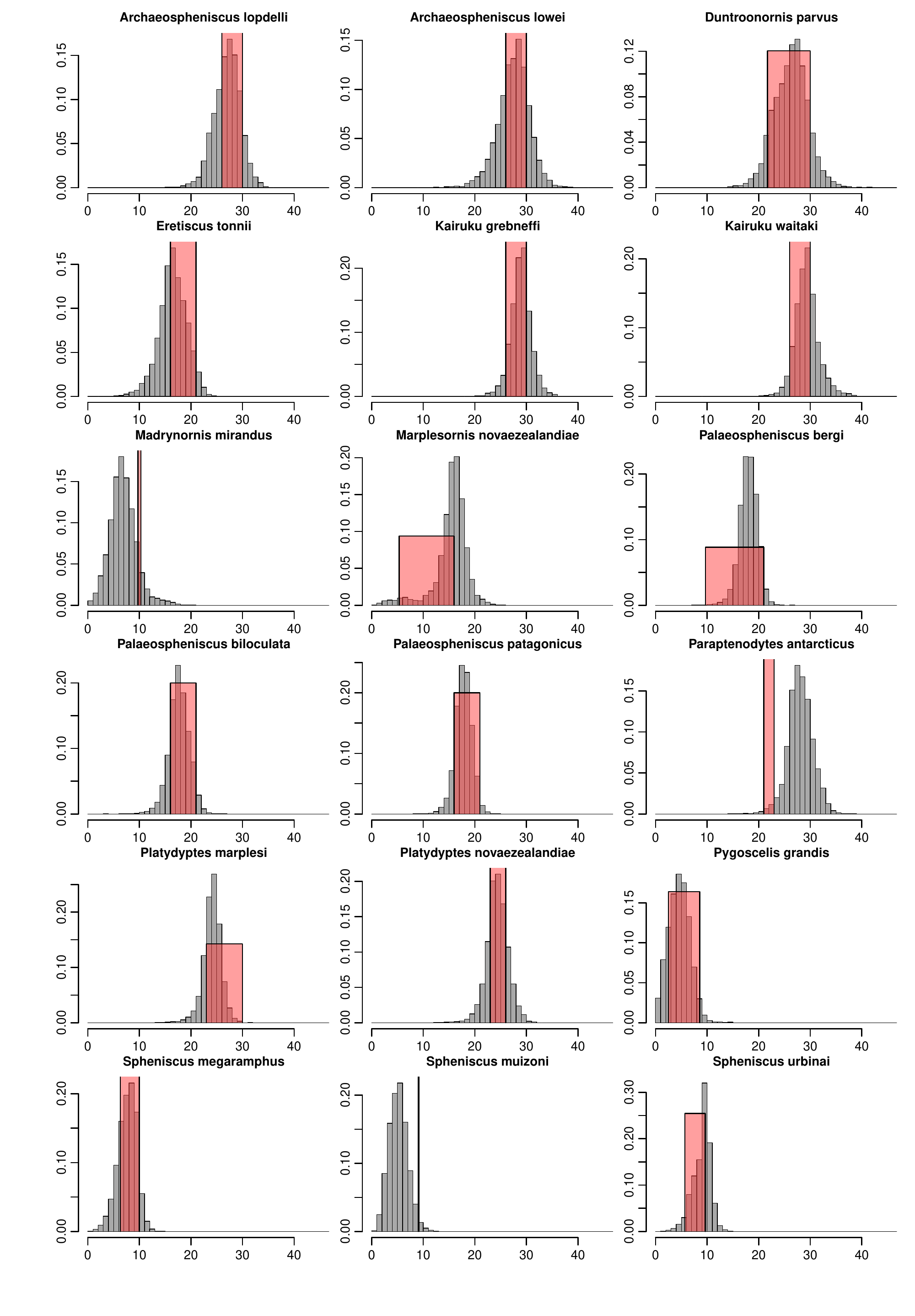}
\caption{\label{hist8_younger}Marginal posterior density plots for the phylogenetic estimate of fossil age of each of the 18 penguin fossils younger than 30 Myr using \Mrelaxed{}. Red boxes are the superimposed age ranges derived from geological data.}
\end{figure}

\begin{figure}
\includegraphics[width=12cm]{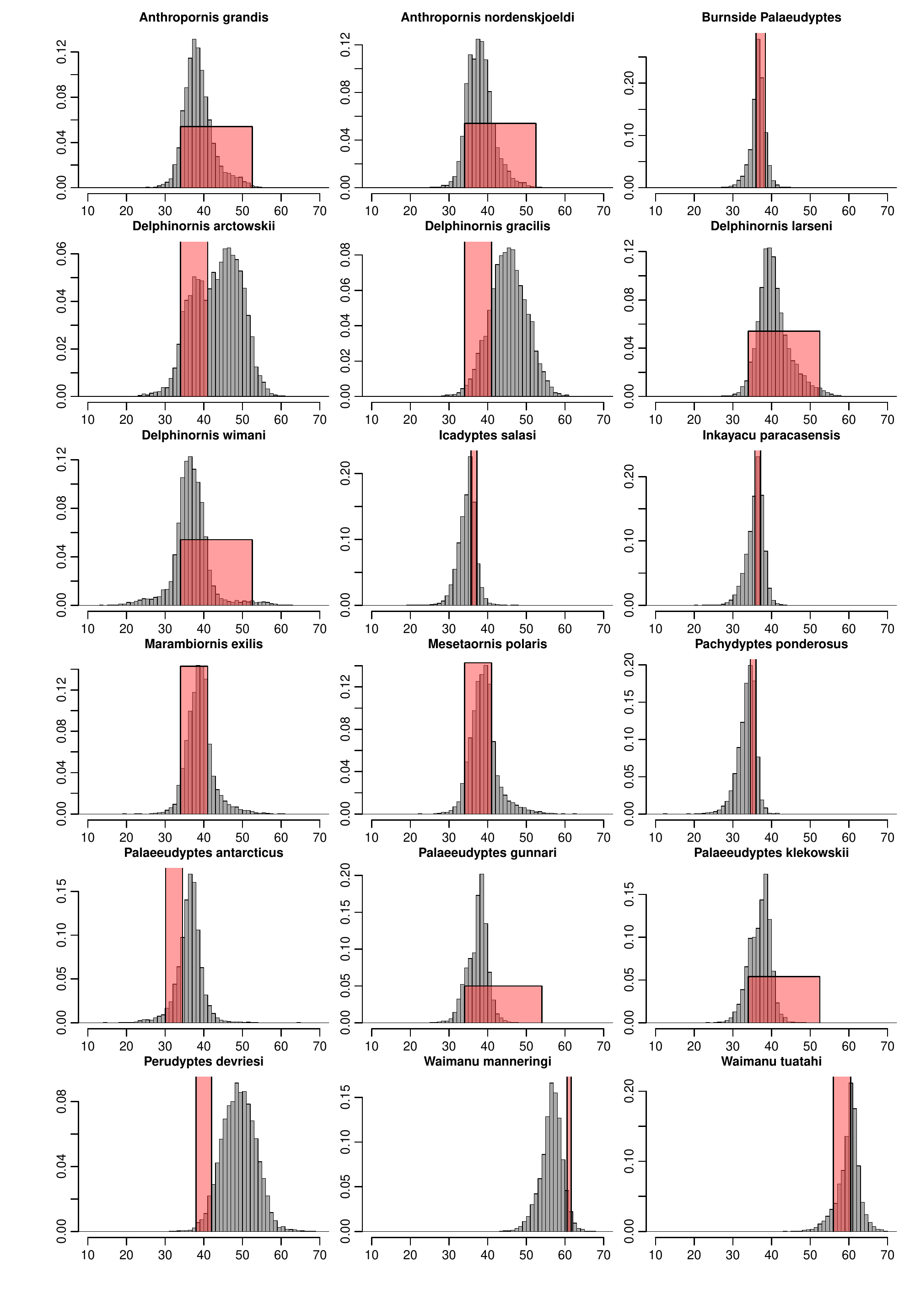}
\caption{\label{hist8_older}Marginal posterior density plots for the phylogenetic estimate of fossil age of each of the 18 penguin fossils older than 30 Myr using \Mrelaxed{}. Red boxes are the superimposed age ranges derived from geological data.}
\end{figure}

\subsection*{Comparison of simple and complex model results}

\begin{figure}
\includegraphics{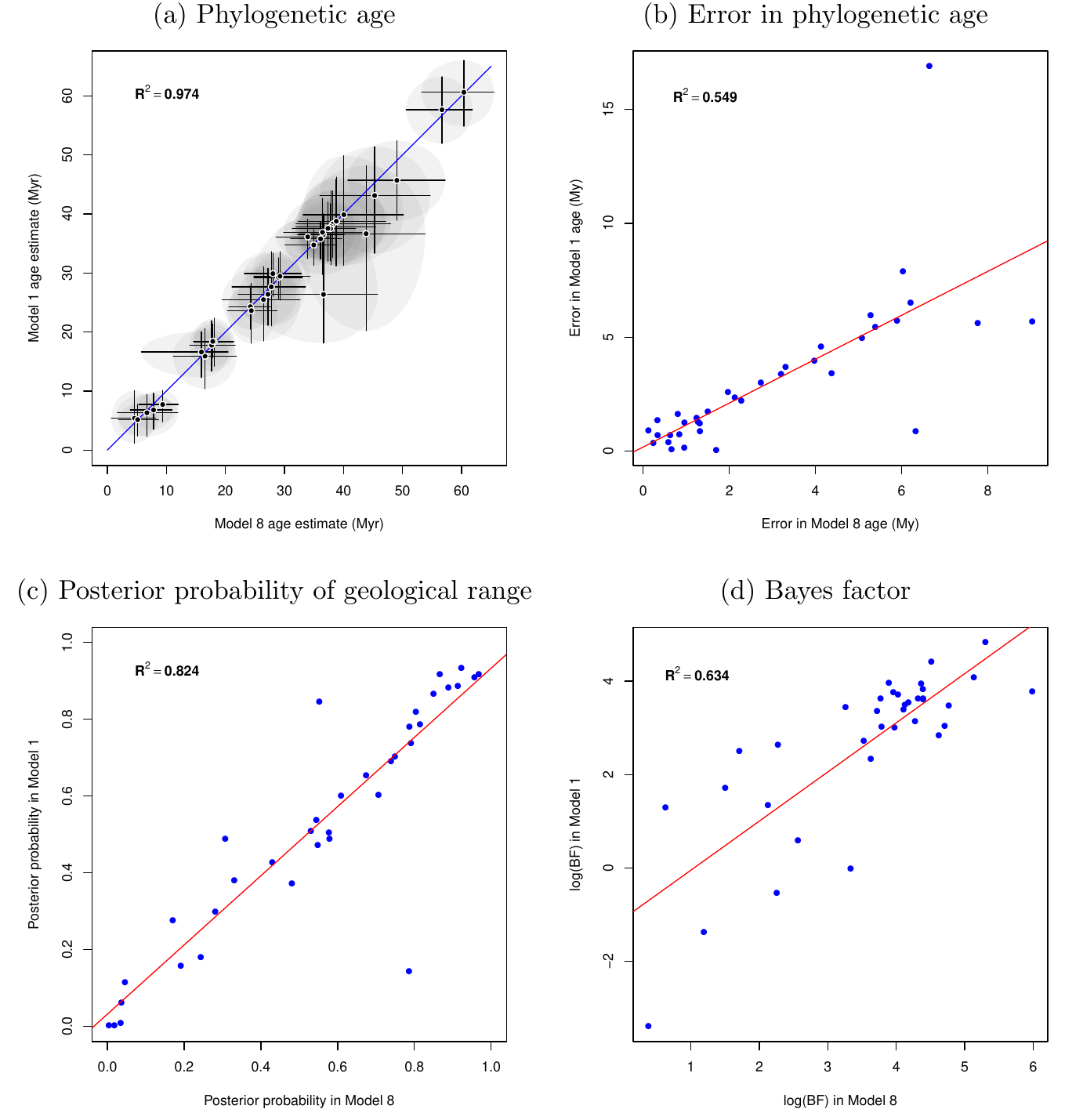}
\caption{\label{fig:compareM1M8}
Comparison of \Mstrict{} against \Mrelaxed{} analyses for the penguin data set. (a) Phylogenetic estimate of fossil age of \Mstrict{} against \Mrelaxed{} with $x=y$ line (blue), (b) Regression of error in phylogenetic estimate of fossil age of \Mstrict{} against \Mrelaxed{}, (c) Regression of posterior probability of palaeontological range of \Mstrict{} against \Mrelaxed{}, (d) Regression of Bayes factor (BF) for palaeontological range of \Mstrict{} against \Mrelaxed{}.}
\end{figure}

Overall the results of analyzing the penguin data set with the \Mstrict{} and \Mrelaxed{} models were strikingly concordant. 
Figure \ref{fig:compareM1M8} shows four regressions between the two models: (a) Regression of estimated phylogenetic estimates of fossil ages of \Mstrict{} against \Mrelaxed{}, (b) Regression of the error in the phylogenetic estimates of fossil ages of \Mstrict{} against \Mrelaxed{} (c) Regression of posterior probability of palaeontological range of \Mstrict{} against \Mrelaxed{}, (d) Regression of Bayes factor (BF) for palaeontological range of \Mstrict{} against \Mrelaxed{}. 
%TS is the following quantified? otherwise I'd delete the next sentence!?
Under \Mrelaxed{} all fossils have positive evidence for their geological age range, whereas under \Mstrict{} there are a handful of fossils with negative evidence for the corresponding geological age range. Furthermore, assuming the median geological age is the truth, the variance in the phylogenetic estimation error of the fossil ages is larger under \Mstrict{} than under \Mrelaxed{}.
%AJD added
This evidence, along with the previous result that \Mrelaxed{} has a higher marginal likelihood than \Mstrict{} \cite{gavryushkina2015bayesian} suggests that the relaxed model is overall a better fit to the data. 
Under both models, there is a positive correlation between the precision of the age estimate and the number of non-ambiguous characters coded for the fossil taxon (Figure \ref{fig:precisionVsnumberCharacters}).

% latex table generated in R 3.0.3 by xtable 1.7-4 package
% Thu Oct 29 11:14:00 2015
\begin{table}[ht]
\centering
\scriptsize
\begin{tabular}{rrrrrrrr}
  \hline
 & post & BF & phylo age & lower & upper & error & ESS \\ 
  \hline
Anthropornis grandis & 0.93 & 82.30 & 38.3 & 32.6 & 43.9 & 4.96 & 2118 \\ 
  Anthropornis nordenskjoeldi & 0.89 & 45.82 & 37.8 & 31.4 & 43.9 & 5.44 & 2694 \\ 
  Archaeospheniscus lopdelli & 0.50 & 20.20 & 26.4 & 21.1 & 30.8 & 1.61 & 2500 \\ 
  Archaeospheniscus lowei & 0.51 & 20.49 & 27.6 & 21.1 & 33.6 & 0.35 & 2076 \\ 
  Burnside Palaeudyptes & 0.54 & 51.19 & 36.5 & 33.2 & 39.8 & 0.69 & 2039 \\ 
  Delphinornis arctowskii & 0.49 & 13.95 & 36.6 & 20.2 & 48.2 & 0.86 & 328 \\ 
  Delphinornis gracilis & 0.27 & 5.54 & 43.1 & 33.3 & 51.4 & 5.61 & 1092 \\ 
  Delphinornis larseni & 0.91 & 58.87 & 39.9 & 31.3 & 49.8 & 3.38 & 955 \\ 
  Delphinornis wimani & 0.14 & 0.98 & 26.4 & 18.1 & 40.0 & 16.89 & 207 \\ 
  Duntroonornis parvus & 0.82 & 37.62 & 25.5 & 18.5 & 31.1 & 0.37 & 2113 \\ 
  Eretiscus tonnii & 0.47 & 10.24 & 15.9 & 10.4 & 20.5 & 2.59 & 3350 \\ 
  Icadyptes salasi & 0.18 & 15.19 & 34.7 & 31.3 & 37.6 & 1.71 & 2931 \\ 
  Inkayacu paracasensis & 0.38 & 42.57 & 35.8 & 32.3 & 38.6 & 0.68 & 3215 \\ 
  Kairuku grebneffi & 0.65 & 37.25 & 29.3 & 25.6 & 32.5 & 1.25 & 3506 \\ 
  Kairuku waitaki & 0.60 & 29.66 & 29.4 & 25.4 & 33.6 & 1.44 & 2653 \\ 
  Madrynornis mirandus & 0.01 & 0.58 & 6.3 & 2.3 & 9.4 & 3.68 & 2581 \\ 
  Marambiornis exilis & 0.70 & 34.56 & 38.7 & 31.8 & 46.2 & 1.20 & 3868 \\ 
  Marplesornis novaezealandiae & 0.37 & 1.81 & 16.6 & 12.3 & 20.0 & 5.96 & 3716 \\ 
  Mesetaornis polaris & 0.69 & 32.69 & 38.8 & 31.2 & 45.9 & 1.26 & 3963 \\ 
  Pachydyptes ponderosus & 0.30 & 28.76 & 36.1 & 32.4 & 39.1 & 0.86 & 2916 \\ 
  Palaeeudyptes antarcticus & 0.16 & 3.79 & 36.9 & 29.7 & 42.6 & 4.58 & 1499 \\ 
  Palaeeudyptes gunnari & 0.88 & 40.55 & 37.5 & 32.1 & 41.7 & 6.50 & 906 \\ 
  Palaeeudyptes klekowskii & 0.86 & 37.53 & 37.5 & 32.0 & 42.0 & 5.71 & 724 \\ 
  Palaeospheniscus bergi & 0.92 & 43.18 & 18.3 & 14.2 & 22.3 & 2.99 & 1243 \\ 
  Palaeospheniscus biloculata & 0.74 & 32.16 & 17.8 & 13.4 & 21.9 & 0.73 & 1186 \\ 
  Palaeospheniscus patagonicus & 0.92 & 124.83 & 18.4 & 15.6 & 21.4 & 0.09 & 1366 \\ 
  Paraptenodytes antarcticus & 0.00 & 0.03 & 29.9 & 26.1 & 33.3 & 7.89 & 2140 \\ 
  Perudyptes devriesi & 0.11 & 3.66 & 45.7 & 38.9 & 52.4 & 5.68 & 1528 \\ 
  Platydyptes marplesi & 0.78 & 36.73 & 24.3 & 20.5 & 27.8 & 2.21 & 4507 \\ 
  Platydyptes novaezealandiae & 0.49 & 22.80 & 23.6 & 18.1 & 28.2 & 0.88 & 6141 \\ 
  Pygoscelis grandis & 0.79 & 20.88 & 5.4 & 1.2 & 10.1 & 0.15 & 2160 \\ 
  Spheniscus megaramphus & 0.60 & 16.86 & 6.8 & 3.5 & 9.6 & 1.36 & 1278 \\ 
  Spheniscus muizoni & 0.00 & 0.25 & 5.2 & 2.4 & 7.7 & 3.95 & 5732 \\ 
  Spheniscus urbinai & 0.85 & 52.29 & 7.7 & 4.8 & 10.1 & 0.05 & 1033 \\ 
  Waimanu manneringi & 0.06 & 12.17 & 57.6 & 51.9 & 63.2 & 3.42 & 4185 \\ 
  Waimanu tuatahi & 0.43 & 31.03 & 60.6 & 54.8 & 66.0 & 2.34 & 3867 \\ 
   \hline
\end{tabular}
\caption{Summary of results for 36 fossil penguins under Model 1. {\em post} is the posterior probability that the phylogenetic age is within the palaeontological age range. {\em BF} is the bayes factor in support of the palaeontological age. {\em phylo age} is the phylogenetic estimate of the age, along with the upper and lower of the corresponding 95\% HPD credible interval. {\em error} is the difference in millions of years between the phylogenetic point estimate of the fossil's age and the mean of it's palaeontological age range. {\em ESS} is the estimated effective sample size for the phylogenetic age estimate.} 
\label{fossilTable1}
\end{table}

% latex table generated in R 3.0.3 by xtable 1.7-4 package
% Fri Oct 30 10:26:45 2015
\begin{table}[ht]
\centering
\footnotesize
\begin{tabular}{rrrrrrrr}
  \hline
 & post & BF & phylo age & lower & upper & error & ESS \\ 
  \hline
Anthropornis grandis & 0.92 & 69.90 & 38.2 & 31.9 & 48.0 & 5.08 & 234 \\ 
  Anthropornis nordenskjoeldi & 0.91 & 61.86 & 37.9 & 31.9 & 45.5 & 5.39 & 302 \\ 
  Archaeospheniscus lopdelli & 0.58 & 27.05 & 27.2 & 22.1 & 32.0 & 0.81 & 1227 \\ 
  Archaeospheniscus lowei & 0.53 & 22.35 & 27.8 & 21.1 & 33.5 & 0.24 & 1134 \\ 
  Burnside Palaeudyptes & 0.54 & 52.95 & 36.6 & 32.2 & 40.1 & 0.63 & 699 \\ 
  Delphinornis arctowskii & 0.31 & 6.49 & 43.8 & 32.5 & 53.8 & 6.33 & 122 \\ 
  Delphinornis gracilis & 0.17 & 3.01 & 45.3 & 36.0 & 54.7 & 7.76 & 421 \\ 
  Delphinornis larseni & 0.96 & 130.06 & 40.1 & 33.2 & 50.1 & 3.20 & 470 \\ 
  Delphinornis wimani & 0.79 & 21.56 & 36.6 & 23.9 & 45.7 & 6.64 & 167 \\ 
  Duntroonornis parvus & 0.80 & 34.45 & 26.4 & 19.5 & 32.6 & 0.59 & 524 \\ 
  Eretiscus tonnii & 0.55 & 13.92 & 16.5 & 11.1 & 21.9 & 1.97 & 851 \\ 
  Icadyptes salasi & 0.24 & 22.50 & 35.0 & 30.1 & 38.8 & 1.50 & 680 \\ 
  Inkayacu paracasensis & 0.33 & 34.59 & 36.1 & 31.0 & 39.6 & 0.34 & 757 \\ 
  Kairuku grebneffi & 0.67 & 41.06 & 29.0 & 24.8 & 33.0 & 0.96 & 2022 \\ 
  Kairuku waitaki & 0.61 & 30.82 & 29.2 & 24.8 & 34.3 & 1.24 & 1233 \\ 
  Madrynornis mirandus & 0.03 & 2.81 & 6.7 & 1.7 & 11.9 & 3.31 & 626 \\ 
  Marambiornis exilis & 0.75 & 43.68 & 38.8 & 32.6 & 47.1 & 1.32 & 639 \\ 
  Marplesornis novaezealandiae & 0.48 & 2.83 & 15.9 & 5.8 & 20.4 & 5.28 & 333 \\ 
  Mesetaornis polaris & 0.74 & 41.45 & 38.8 & 32.3 & 47.1 & 1.27 & 514 \\ 
  Pachydyptes ponderosus & 0.28 & 26.40 & 33.9 & 28.6 & 37.6 & 1.32 & 1438 \\ 
  Palaeeudyptes antarcticus & 0.19 & 4.85 & 36.4 & 29.9 & 43.1 & 4.13 & 333 \\ 
  Palaeeudyptes gunnari & 0.89 & 43.82 & 37.8 & 32.1 & 42.0 & 6.21 & 269 \\ 
  Palaeeudyptes klekowskii & 0.85 & 33.30 & 37.4 & 31.3 & 41.9 & 5.89 & 387 \\ 
  Palaeospheniscus bergi & 0.97 & 120.25 & 18.1 & 14.7 & 21.4 & 2.73 & 596 \\ 
  Palaeospheniscus biloculata & 0.79 & 43.49 & 17.7 & 13.9 & 21.6 & 0.84 & 595 \\ 
  Palaeospheniscus patagonicus & 0.87 & 74.30 & 17.8 & 14.6 & 21.0 & 0.66 & 565 \\ 
  Paraptenodytes antarcticus & 0.02 & 0.63 & 28.0 & 23.2 & 32.8 & 6.03 & 1238 \\ 
  Perudyptes devriesi & 0.05 & 1.37 & 49.0 & 40.7 & 57.2 & 9.03 & 432 \\ 
  Platydyptes marplesi & 0.79 & 38.30 & 24.2 & 20.6 & 27.6 & 2.28 & 1704 \\ 
  Platydyptes novaezealandiae & 0.58 & 32.98 & 24.4 & 20.3 & 28.7 & 0.13 & 1530 \\ 
  Pygoscelis grandis & 0.81 & 25.11 & 4.6 & 0.7 & 8.2 & 0.96 & 1631 \\ 
  Spheniscus megaramphus & 0.71 & 26.71 & 7.8 & 3.9 & 11.0 & 0.33 & 753 \\ 
  Spheniscus muizoni & 0.00 & 0.95 & 5.1 & 1.9 & 8.7 & 3.98 & 1574 \\ 
  Spheniscus urbinai & 0.55 & 11.77 & 9.4 & 5.3 & 12.0 & 1.70 & 586 \\ 
  Waimanu manneringi & 0.04 & 7.16 & 56.7 & 50.6 & 61.8 & 4.38 & 1939 \\ 
  Waimanu tuatahi & 0.43 & 31.52 & 60.4 & 53.2 & 65.5 & 2.13 & 2157 \\ 
   \hline
\end{tabular}
\caption{Summary of results for 36 fossil penguins under Model 8. {\em post} is the posterior probability that the phylogenetic age is within the palaeontological age range. {\em BF} is the bayes factor in support of the palaeontological age. {\em phylo age} is the phylogenetic estimate of the age, along with the upper and lower of the corresponding 95\% HPD credible interval. {\em error} is the difference in millions of years between the phylogenetic point estimate of the fossil's age and the mean of it's palaeontological age range. {\em ESS} is the estimated effective sample size for the phylogenetic age estimate.} 
\label{fossilTable8}
\end{table}

\subsection*{Canids conform well to a morphological clock}

\begin{figure}
\includegraphics{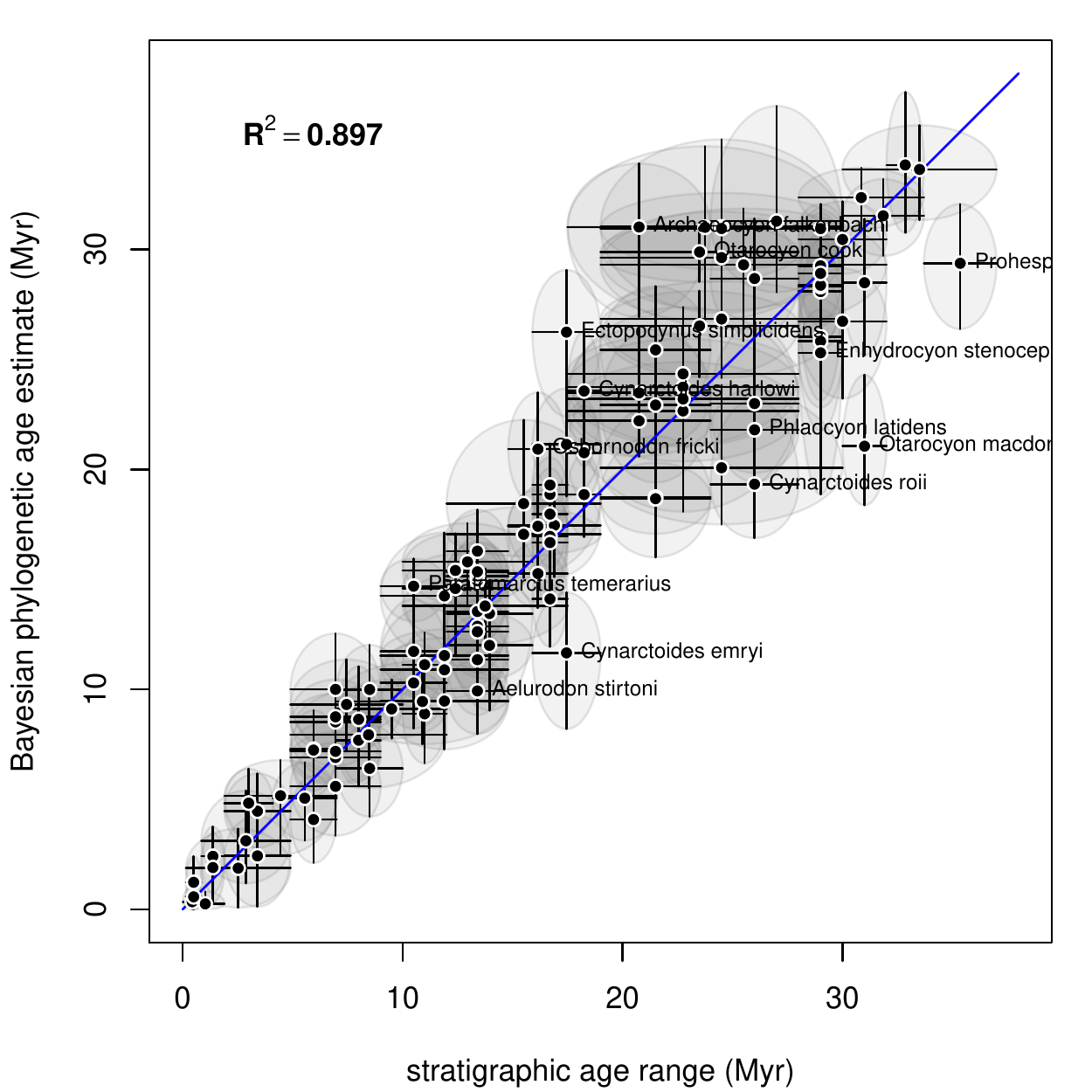}
\caption{\label{fig:canidMorphVsGeo}
The Bayesian phylogenetic estimate of fossil age (median and 95\% credible interval of marginal posterior) for each of the \ncanidfossils{} canid fossils plotted against their stratigraphic age ranges, under a strict morphological clock model \Mstrict{}. 
%The Bayesian estimates are represented by the median of the marginal posterior distribution and the upper and lower limits of the 95\% HPD interval. 
The palaeontological age estimates are represented by the mid-point of the range and the upper and lower limits. The Bayesian esti- mates are represented by the median of the marginal posterior distribution and the upper and lower limits of the 95\% HPD interval.
Blue line shows the $x=y$. 
If the vertical line doesn't cross $x=y$, then the mean of the stratigraphic age range is not in the credible interval of the phylogenetic estimate of fossil age. The 13 fossils for which this is the case are labelled.}
%TS I find this plot very hard to read as many fossils are overlaying...can we do something there for better visualization?
\end{figure}

The canid data set shows remarkable consistency between stratigraphic age ranges and phylogenetic estimates of fossil ages, even with the simple strict morphological clock model (\Mstrict{}).
The $R^2 = 0.897$ between the phylogenetic and stratigraphic ages ranges (see Figure \ref{fig:canidMorphVsGeo}). 
Only 13 out of \ncanidfossils{} fossils (11\%) don't have the mean stratigraphic age in the 95\% credible interval of the phylogenetic estimate of fossil age and there are no extreme outliers. 
The median error is $1.56$ My, which in absolute terms is more accurate than the age estimates for the penguin data set. However the median relative error was 13.2\%, more than twice that for the penguin fossils. 

This data set contains half as many morphological characters as does the penguin data set ($122$ versus $245$), nevertheless the individual age estimates are much more precise in absolute terms (median HPD range = 4.2 Myr for canids as opposed to 9.6 Myr for penguins). However this is mainly due to the fact that the average age of the penguin fossils is considerably larger and the median relative precision (i.e. RSD) was 7.2\%, only slightly better than the value for the penguin fossils of 9.2\%.

\begin{figure}
\includegraphics[width=15cm]{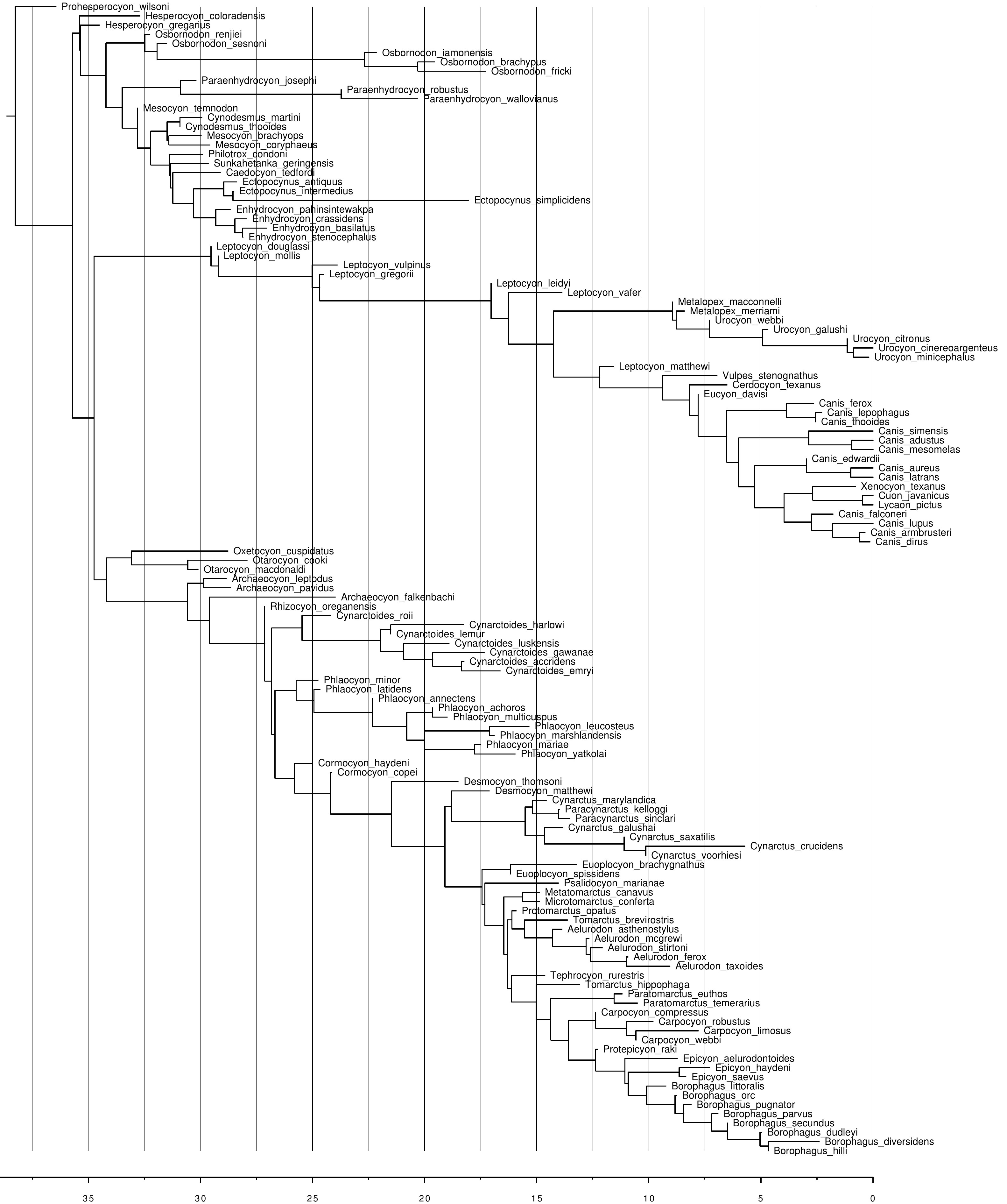}
\caption{\label{fig:canidTree} A sample from the posterior distribution of an analysis of the canid data set, 
%\gslater{%
showing three main clades, one clade with extant representatives and two extinct clades ({\it Hesperocyoninae} and {\it Borophaginae}). The x-axis is in units of Mya. }
%}%
\end{figure}

Figure \ref{fig:canidTree} shows a sample from the posterior distribution of the analysis of the canid data set. The tree has three main clades, one clade with extant representatives and two extinct clades ({\it Hesperocyoninae} and {\it Borophaginae}).

Table \ref{estimatesTable} shows that the rate of morphological evolution in canids is faster than that estimated in the penguins, however this could be a simple reflection of the shorter geological time scale (and shorter average branch lengths) over which the rate has been estimated \cite{Gingerich1993}.

\begin{table}[ht]
\centering
\begin{tabular}{c|rrr}
  \hline
Analysis & $t_{MRCA}$ & morph. rate  & log sd rate (S) \\ 
 & (Myr) & ($\times 10^{-2}$/Myr) & \\
  \hline
Penguins Mk-1 & 61.7 [60.5, 63.8] & 1.79 [1.53, 2.05] & -  \\
Penguins Mk-8 & 61.4 [60.5, 63.3] & 1.29 [0.77, 1.90] & 0.69 [0.40, 0.99] \\
  \hline
Canids Mk-1 &  36.8 [35.4, 38.5] & 2.83 [2.47, 3.19] & - \\
\hline
  \end{tabular}
\caption{Summary of key parameters for the three main analyses. Note: the $t_{MRCA}$ is the time of the most recent common ancestor of all taxa, including both extinct and extant species. The 95\% HPD interval for each estimate is in square brackets. The morphological clock rate is given in percent change per million years.} 
\label{estimatesTable}
\end{table}

\section*{Discussion}

In this paper we have demonstrated that even a small number of morphological characters (some of the fossils had as few as 7 morphological traits coded) can be used in the context of a rich fossil reference data set to provide an accurate age of the fossil based on a phylogenetic model. 
In all cases we used the new fossilized birth-death tree prior, which is a crucial ingredient in allowing for the estimating of fossil ages under a birth-death tree prior. 

%AJD
We found that although a strict morphological clock does a surprisingly good job of estimating fossil ages, there is evidence that phylogenetic estimation of penguin fossil ages is improved by a model that includes a variation in rates of morphological evolution among lineages. 
However, in the penguin data set the variation in evolutionary rates was not too extreme and the estimated log standard deviation of the relaxed morphological clock ($S = 0.69$; refer to Table \ref{estimatesTable}) is comparable to values obtained for molecular clocks.   
The median error in age estimates for the two data sets investigated were 2 My and 1.6 My respectively, using either a very simple or more complex models of discrete morphological change. 
%%%

In absolute terms the fossil estimates were both slightly more accurate and more precise on average in the canid data set. 
One might think that the larger reference set of fossils in the canid data set (115 versus 35) makes up for the smaller number of characters (122 versus 245) with regards to accuracy and precision of fossil age estimates. 
However, since the average age of the canid fossils is considerably younger than that for the penguin fossils, a more appropriate comparison uses relative error and relative precision. 
By these measures the penguin data set actually provides the more accurate estimates, whereas relative precision is overall slightly better for the canid data set. Future work is needed to investigate in a more systematic fashion how the amount of morphological data available for a new fossil and the number of related reference fossils of known age affect the accuracy and precision of the phylogenetic estimate of a fossil's age.

%\gslater{%
Another difference between the two data sets analyzed here is that the penguin fossils were largely single specimens, or at least single localities, so that the age range specified for the fossil represents uncertainty in the geological age of the horizon the fossil was associated with (for example, uncertainty in radiometric dates from the volcanic layers above or below the fossil-carrying horizon and uncertainty about the age difference between the volcanic layers and the horizon the fossil is in). 
On the other hand, most of the canid species were assigned stratigraphic age ranges based on multiple specimens from multiple localities spanning a substantial time range. 
For example, there are thousands of specimens of {\it Hesperocyon gregarious} from multiple sites in North America spanning $>5$ Myr (pers. comm. Graham J. Slater). 
In this context it is interesting to note the canids that fall off the x=y line in Figure 7 are mostly (but not exclusively) taxa represented by singletons and therefore those with relatively short stratigraphic ranges. 
This raises the question of whether multiple specimens of a single species that span a significant time frame and/or different localities should be coded as separate taxa as input for the fossilized birth-death method. Even if not coded as separate taxa it may be possible to extend the method used here to explicitly account for multiple specimens and associated ages when a fossil species is represented by more than one fossil. We leave these considerations for future work. 
%}%

There are diverse potential applications for this methodology. 
The most obvious is the estimation of dates for fossils that are temporally unconstrained, either due to poor knowledge of the age of the sediments in which it was found, 
%\alexei{%
or because the fossil was not associated with a horizon of known age, e.g. \cite{Berger2015},
%}%
or because of a complete lack of provenance data (e.g. a recent fossil described as a `four-legged snake' has excited controversy for a lack of provenance\footnote{See \url{http://news.sciencemag.org/paleontology/2015/07/four-legged-snake-fossil-stuns-scientists-and-ignites-controversy}}). 
It can also be used a way of testing the ``morphological clock'' and to discover potential problems in the data by identifying outlier fossils with respect to model fit. 
Overall, we anticipate that this approach will help to promote the application of a consistent probabilistic framework to consider both molecular and fossil evidence. Our results are encouraging in suggesting that the statistical models presented are adequate for inference of phylogenetic time-trees from morphological fossil data.

\section*{Availability}

All BEAST2 xml input files and R analysis scripts required to reproduce the results in this paper are available at \url{https://github.com/alexeid/fossilDating}.

\section*{Acknowledgements}

The authors would like to thank Daniel Ksepka and Graeme T. Lloyd for insightful comments and suggestions on early drafts of this manuscript. %\alexei{%
We are also very grateful to Graham J. Slater for comments on the phylogenetic analysis of canids and for providing stratigraphic age ranges for the canid fossils. We thank Mario dos Reis and one anonymous reviewer for very helpful comments during revision.
%}%
AJD and TS were partially funded by Marsden grant UOA1324 from the Royal Society of New Zealand (\url{http://www.royalsociety.org.nz/programme?s/funds/marsden/awards/2013-awards/}). AJD was funded by a Rutherford Discovery Fellowship from the Royal Society of New Zealand (\url{http://www.royalsociety.org.nz}).  TS was supported in part by the European Research Council under the 7th Framework Programme of the European Commission (PhyPD; grant agreement 335529). AJD and TS also thank ETH Z\"{u}rich for funding.

\bibliographystyle{vancouver}

\bibliography{fossilDating}

\end{document}